\documentclass[letterpaper, 10 pt, journal, twoside]{IEEEtran}
\usepackage{xcolor,soul,framed} 
\usepackage{lineno}
\usepackage{fancyhdr}
\colorlet{shadecolor}{yellow}
\usepackage[pdftex]{graphicx}
\graphicspath{{../pdf/}{../jpeg/}}
\DeclareGraphicsExtensions{.pdf,.jpeg,.png}

\usepackage[cmex10]{amsmath}
\usepackage{array}
\usepackage{mdwmath}
\usepackage{mdwtab}
\usepackage{eqparbox}
\usepackage{url}

\usepackage{caption}
\usepackage{subcaption}
\usepackage{graphicx}
\usepackage[export]{adjustbox}
\usepackage{authblk}
\usepackage{gensymb}
\IEEEoverridecommandlockouts
\usepackage{titlesec}

\titlespacing*{\section}{0pt}{1.7em}{0.5em} 
\titlespacing*{\subsection}{0pt}{1.7em}{0.5em} 

\AtBeginDocument{
  \setlength{\intextsep}{-1pt plus 1pt minus 1pt}

  \setlength{\textfloatsep}{-1pt plus 1pt minus 1pt}

  \setlength{\floatsep}{-1pt plus 1pt minus 1pt}
}

\fancypagestyle{firstpage}{
  \fancyhf{} 
  \fancyhead[C]{\scriptsize This article has been accepted for publication in IEEE Robotics and Automation Letters (RA-L). The final published version is available at: \url{https://doi.org/ 10.1109/LRA.2024.3514513}}
  \fancyfoot[C]{\scriptsize © 2025 IEEE. Personal use of this material is permitted.  Permission from IEEE must be obtained for all other uses, in any current or future media, including reprinting/republishing this material for advertising or promotional purposes, creating new collective works, for resale or redistribution to servers or lists, or reuse of any copyrighted component of this work in other works.}
}

\begin{document}
\title{H-Net: A Multitask Architecture for Simultaneous 3D Force Estimation and Stereo Semantic Segmentation in Intracardiac Catheters}

\author{Pedram Fekri$^{1}$,
            Mehrdad Zadeh$^{2}$,
            Javad Dargahi$^{1}$

\thanks{This article has been accepted for publication in IEEE Robotics and Automation Letters (RA-L). The final published version is available at: \protect\url{https://doi.org/10.1109/LA.2024.3514513}
} 
\thanks{$^{1}$Pedram Fekri and Javad Dargahi are with the Mechanical, Industrial and Aerospace Engineering Department, Concordia University, Montréal, Quebec, Canada
        {\tt\footnotesize p$\_$fekri@encs.concordia.ca, dargahi@encs.concordia.ca}}%

\thanks{$^{2} $Mehrdad Zadeh is with the Electrical and Computer Engineering Department, Kettering University, Flint, Michigan, USA
        {\tt\footnotesize mzadeh@kettering.edu}}

}


\maketitle

\thispagestyle{firstpage} 
\pagestyle{plain}
\begin{abstract}
The success rate of catheterization procedures is closely linked to the sensory data provided to the surgeon.
Vision-based deep learning models can deliver both tactile and visual information in a sensor-free manner, while also being cost-effective to produce. Given the complexity of these models for devices with limited computational resources, research has focused on force estimation and catheter segmentation separately. 
However, there is a lack of a comprehensive architecture capable of simultaneously segmenting the catheter from two different angles and estimating the applied forces in 3D. 
To bridge this gap, this work proposes a novel, lightweight, multi-input, multi-output encoder-decoder-based architecture. It is designed to segment the catheter from two points of view and concurrently measure the applied forces in the $x$, $y$, and $z$ directions. This network processes two simultaneous X-Ray images, intended to be fed by a biplane fluoroscopy system, showing a catheter's deflection from different angles. It uses two parallel sub-networks with shared parameters to output two segmentation maps corresponding to the inputs. Additionally, it leverages stereo vision to estimate the applied forces at the catheter's tip in 3D. The architecture features two input channels, two classification heads for segmentation, and a regression head for force estimation through a single end-to-end architecture. The output of all heads was assessed and compared with the literature, demonstrating state-of-the-art performance in both segmentation and force estimation. To the best of the authors' knowledge, this is the first time such a model has been proposed.

\vspace{2mm}
\noindent DOI: \url{https://doi.org/ 10.1109/LRA.2024.3514513}
\end{abstract}


%
\IEEEpeerreviewmaketitle


\section{Introduction}

In a cardiac catheterization procedure, a surgeon explores the cardiovascular system for various purposes, including diagnostic testing, measuring pressure and oxygen levels, conducting biopsies, or delivering treatment. This is achieved by steering a long, flexible tube known as a catheter into the patient's vascular system, typically through the groin, neck, or shoulder, guided by an X-Ray fluoroscopy imaging system 
\cite{complic:30285356, complic2}. 
While the efficacy of interventional catheterization treatments is well-documented, the procedure is not without its imperfections and safety concerns. These issues primarily fall into two categories: the lack of tactile feedback and challenges in catheter localization.
Currently, conventional catheters do not offer surgeons the ability to feel or sense of touch when the catheter's tip interacts with anatomical structures such as heart or vessel tissues. For instance, during an ablation procedure, the surgeon must apply a precise force between 0.1N and 0.3N at the catheter's tip against the targeted defective spots.  
Moreover, inserting and maneuvering the catheter can result in unintended movements due to the complex vascular structures and flexibility of both the catheter and vessels. Without enhanced visualization, the risk of causing tears or abrasions, leading to life-threatening bleeding, increases significantly. Enhanced visualization minimizes these risks and ensures precise navigation and accurate tissue measurement, improving procedural safety and outcomes.
Having an accurate visualization of the catheter is also crucial for developing autonomous or semi-autonomous robotic systems for catheterization 
\cite{lo2021safety,chen2020safety,fekri1, fekri2, shah2015real}.
\par

\par Regarding the measurement of applied forces, sensor-free approaches mentioned in the literature aim to reduce the production costs of catheters with embedded sensors (e.g., TactiCath™ contact force ablation catheter by Abbott (Abbott Park, Illinois, United States)). 
For instance, model-based methods extract mechanical features of catheters within the images to estimate the forces \cite{khoshnam2012modeling}. Learning-based models employ deep learning architectures to estimate forces from the deflections. The input for these models can be either extracted features from images of the catheter, simulations, or the images themselves \cite{khodashenas2021, roshanfar2023autonomous, fekri2021deep, roshanfar2023deeplearning}.
However, these solutions face three major challenges: 1) the hand-crafted features are not robust \cite{khodashenas2021}; 2) learning and model-based feature extractors require a segmented shape of the catheter \cite{fekri2021deep}; and 3) the models are incapable of estimating the force in the z-direction \cite{roshanfar2023deeplearning}. These models are designed for deployment on monoplane fluoroscopy, where the catheter's deflection along the z-axis is not trackable in a single 2D image.
However, Y-Net has been proposed to measure forces in 3D by processing two images of the catheter taken by a biplane fluoroscopy \cite{y-net}. 
Although the Y-Net addresses the previously mentioned issue, it still relies on a catheter segmentor in the preprocessing phase to provide a clear shape of the catheter as input, thereby preventing the potential impact of objects other than the distal shaft on the estimation.
\par Furthermore, as a separate task, surgeons need  a visualization that can localize the catheter in X-Ray. This is typically achieved through a separate process using image processing algorithms, such as thresholding or semantic segmentation models
\cite{segmed2, catseg6, segmed_ynet, unet-2024, hrnet_cat, FCN-cat}.
In the majority of catheter segmentation works, U-Net, FCN, or HR-Net is the chosen architecture due to their simplicity. According to the results of these studies, the nature of the data distribution in catheter segmentation is not complex enough to require more sophisticated models. However, its best to consider other CNN and open-set vision transformers-based architectures designed for semantic and instance segmentation \cite{fcn,segnet, maskrcnn, unet, segment-anything}. As pointed, the segmentation model is a requisite tool for both visualization and force estimation in catheterization. Ideally, two deep learning models should run simultaneously to address both needs, though this requires significant computational resources.  
\par 
In response to the aforementioned issues, this work presents a comprehensive solution encompassing catheter force estimation and segmentation for biplane fluoroscopy systems. We propose a novel multi-input, multi-output and multi-task architecture equipped with two parallel sub-networks, each consisting of an encoder and a decoder with the following contributions: This end-to-end architecture is designed to process two X-Ray images of a catheter from two angles simultaneously. By fusing these images, the network can segment the catheter from both perspectives simultaneously through two separate segmentation heads, enabling 3D reconstruction and visualization. Leveraging the feature maps of both decoders to capture the catheter's 3D variation, the force estimation head predicts the applied force at the catheter's tip along the x and y axes, as well as the force along the z axis. Addressing the problem of catheter segmentation in the preprocessing phase, this approach is designed to ensure that the estimation head primarily focuses on the catheter, minimizing the influence of other object variations in the X-ray images.
Although the proposed architecture segments two X-Ray images and estimates forces in 3D simultaneously, the network remains lightweight by sharing parameters between its sub-networks. This balances computational complexity, memory usage, and latency. Additionally, to assess the model's performance on a dataset that closely mimics real-world conditions, we have designed a synthetic X-Ray image generator. This generator superimposes images of actual catheter deflections onto a chest X-Ray background, effectively simulating a catheterization procedure as observed under fluoroscopy. Leveraging this capability, we compiled two separate synthetic X-Ray datasets, each with a different level of difficulty. The proposed model was trained and tested on these two synthetic datasets as well as an additional RGB dataset. Compared to other pioneering methods in the literature, for the first time, the proposed method accomplishes 3D force estimation and stereo segmentation tasks in an end-to-end architecture with superior performance, establishing it as the state-of-the-art. 
\vspace{-10pt}
\section{Multi-Task Encoder-Decoder Network}
The proposed architecture provides the following solutions for the mentioned issues of learning-based models in catheter applied force estimators and segmentation: 1) It directly approximates forces within the X-Ray images, eliminating the need for a separate segmentation engine as a pre-processing step. 2) It is capable of estimating forces in the $x$, $y$, and $z$ directions.
3) it provides a stereo catheter segmentation feature. 
4) Designed as an end-to-end system, it does not significantly increase the computational complexity of the overall setup. 
Inspired by both U-Net and Y-Net, this architecture processes two raw X-Ray images through two parallel encoders with shared parameters \cite{y-net, unet}. The feature maps extracted from both inputs traverses through the decoder and then they are directed toward two segmentation heads and one force estimation head. This graph outputs the segmented shape of the catheter from two perspectives, along with the predicted forces along the $x$, $y$, and $z$ axes. This architecture consists of two parallel sub-networks, each with its own input and output head, connected through a central force estimation head. This configuration resembles the letter 'H'. Hence, for simplicity, we have named it "H-Net". Before delving into the details of H-Net, the following section will describe the process of compiling the requisite dataset for training and evaluating this model.
\vspace{-10pt}
\subsection{Synthetic Data Generation and Data Preparation}
As noted earlier, H-Net is fed by two images of a catheter captured from two angles. Similar to Y-Net, this stereo vision approach allows the network to understand the catheter's depth and 3D deflection variations, enabling force measurement along the $z$ axis. The necessary data embodying these characteristics was generated using a mechanical setup, as detailed in \cite{y-net}. This setup, replicating biplane fluoroscopy, uses two cameras positioned to capture the distal shaft of the catheter from two perspectives. The catheter's tip is manipulated by two motorized linear actuators pressing it against a force sensor, allowing movement in various directions. Each dataset record from this setup comprises two images along with the measured forces in the $x$, $y$, and $z$ directions. Since the Y-Net architecture receives segmented shape of the catheter, the background removes by a thresholding method. Thus, both segmented and original RGB images are available from the previous work. 
\par  

Considering such a configuration, H-Net is trained and evaluated on two types of input images. The first type includes the original unsegmented RGB images compiled in \cite{y-net}. The second type comprises synthetic X-Ray images. Given that the ultimate goal is to deploy such models in real operation rooms using X-Ray data, it is crucial for the H-Net model to be exposed to data closely resembling real-world scenarios. In fact, this approach allows for a more effective comparison with RGB images, which typically have less complex backgrounds than X-Ray images.
To this end, we propose a synthetic data generator in which it utilizes the previously explained data and converts them into synthetic X-Ray images. First, a region containing the aorta, superior vena cava, and heart is cropped from a real chest X-Ray (refer to the background image under the grid Fig. 1). This region is selected as the background of synthetic images. Typically, the pixels representing these organs exhibit higher intensity. This selected region may contain pixels with consistently high intensity (e.g., white areas), resulting in a smoother background and data that are easier to model. However, radiolucent areas, which appear as darker regions, may interfere with the intensity of the catheter, thereby producing more challenging backgrounds. Given such an area, a grid of 16 cells is defined to break down the input region into 16 separate, non-overlapping patches (refer to the patchifier in Fig. 1). These patches are later utilized to create a synthetic background. As depicted in Fig. 1, for each dataset record, the catheter's shape is extracted from the thresholded (segmented) images to serve as the foreground. Subsequently, a blank matrix matching the size of each RGB image is created for the background so that it encompasses the same grid layout described previously. For each image, the previously mentioned patches, derived from the patchifier, are randomly placed in each grid cell of background matrix to construct a background. Finally, as shown in Fig. 1, the foreground is superimposed onto this randomly generated background. This process results in the production of a synthetic X-Ray image and is applied to all samples in the dataset. For both synthetic and RGB dataset, the thresholded version of images are used as the segmentation targets. 

\begin{figure}[t]
	\centering
	\includegraphics[trim={0mm 0mm 0 0},clip, width=0.47\textwidth]{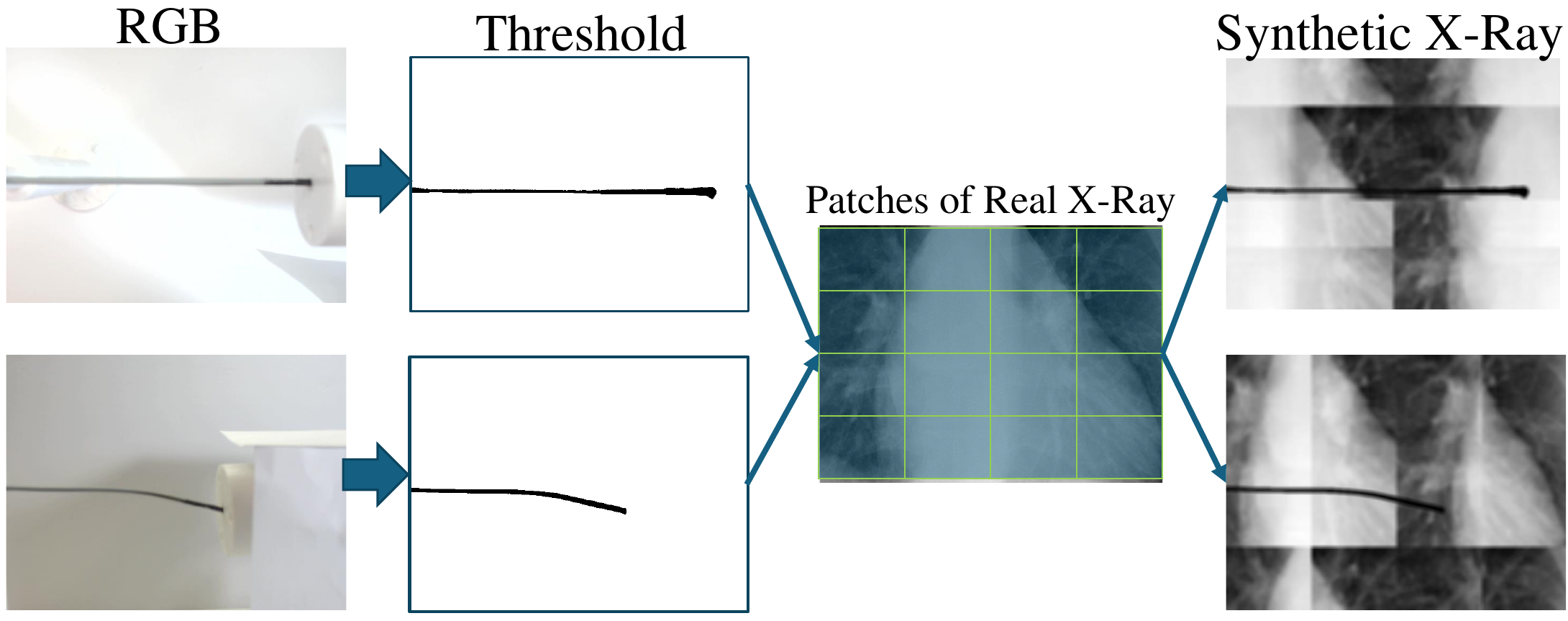}
	\caption{The process of generating synthetic X-Ray images from the RGB and their corresponding thresholded images.}
	\label{fig:synth}
\end{figure}
\begin{figure*}[t]
	
	\centering
	\includegraphics[trim={0mm 0mm 0 0},clip, width=0.72\textwidth]{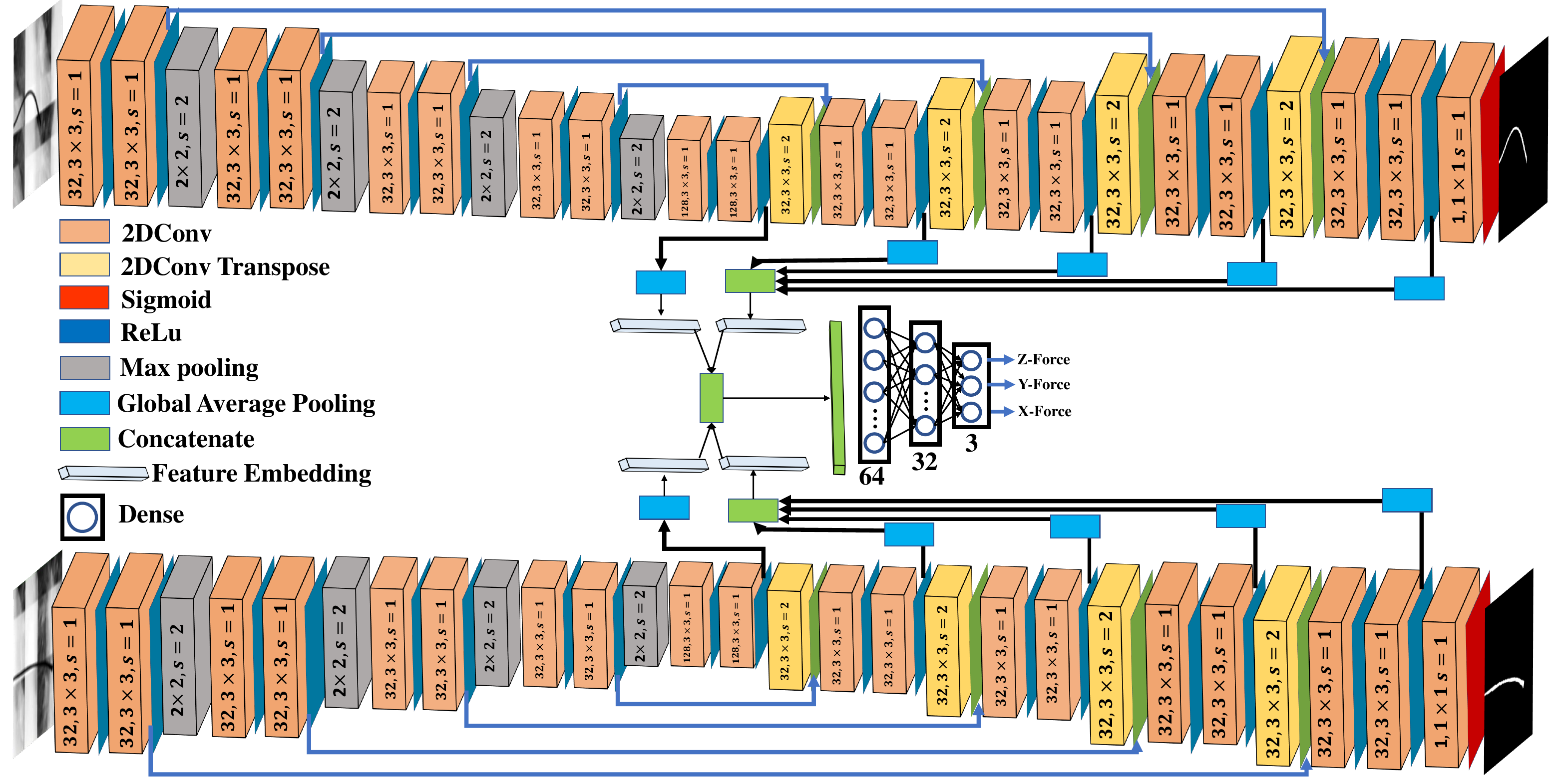}
	\caption{The diagram depicts H-Net's architecture, which includes two sub-networks, each with an encoder and a decoder. All layers of the encoders and decoders are shared between the two sub-networks. The architecture also has two segmentation heads and a central force estimation head. Upon receiving two input images, the network outputs two segmentation maps and the estimated forces in 3D.}
	\label{fig:net}
\end{figure*}
\vspace{-13pt}
\subsection{Methodology}
The tip of a catheter generates forces when it is pressed against the surface of lumens. These forces can be correlated with the deflection shapes of the distal shaft, particularly when the catheter's tendons are not engaged. In practice, the catheter is visualized using continuous X-Ray images provided by a fluoroscopy imaging system. Bearing this in mind, the goal is to segment the catheter and measure the forces generated at the tip directly from these X-Ray images. This problem presents the following challenges: 1) The catheter's distal shaft must be considered in three dimensions, as it can be deflected along any of these axes. 2) Other objects present in the X-Ray images, such as organs, can interfere with the visualization of the catheter's deflection shapes, potentially leading to inaccurate force estimation. 3) Apart from force estimation, catheter visualization can assist surgeons with accurate measurement and steering. 
As mentioned earlier, these three challenges are interdependent but have been addressed independently in \cite{y-net}, \cite{hrnet_cat, catseg6, unet-2024_cat}. However, there currently exists no single network designed to tackle both challenges simultaneously.
\par
The H-Net solves the above-mentioned challenges through a novel end-to-end architecture. The proposed architecture features two encoders and two decoders within two separate sub-networks. As illustrated in Fig. 2, H-Net is a multi-input network capable of processing two X-Ray images simultaneously, providing the network with stereo vision capabilities. Additionally, it is a multi-output network comprising three distinct heads. Two of these heads are dedicated to concurrently segmenting the input X-Ray images, addressing respective classification problems. Utilizing features from both the bottlenecks and the decoders, the third head is designed to estimate forces along the $x$, $y$, and $z$ axes. The parameters of the encoder and decoder are shared between the two sub-networks, with the goal of keeping the model's complexity and memory utilization low. In essence, the lower sub-network is a mirrored version of the upper one. 
The inputs of the network are two X-Ray images of size $I \in R^{(h\times w \times c)}$ showing the catheter's from two viewing angles. The encoder of a sub-network comprises 4 convolutional blocks ($b$). In each block there are two consecutive 2D convolution layers with kernels of $3\times3$ and ReLu activation functions. The output of the second 2D convolution layer is down-scaled by a $2\times2$ max-pooling layer. The encoder extracts higher-level features as the input is directed to the depth of the network. The output of each block is denoted by $\hat{O}^{sn,b}_{enc}$, where $b$ indicates the number of blocks in sub-network $sn$. These feature maps are then sent to the subsequent encoder block, moving towards the bottleneck. Moreover, $\hat{O}^{sn,b}_{copy}$, which is a copy of each encoder block's output before downscaling (max-pooling), is forwarded to the corresponding decoder block within the same sub-network. It's evident that the dimensions $h_b \times w_b$ of $\hat{O}^{sn,b}_{copy}$ match the input size of the decoder block.
In each block of the encoder in both sub-networks, the inputs are down-scaled by a factor of 2, and their feature maps are extracted simultaneously. The subsequent  block after $b=4$ in the encoder, is the bottleneck block. It comprises a stack of two 2D convolution layers with ReLu activation functions. The bottleneck performs two tasks: 1) It supplies the input for the first block of the decoder in each sub-network. 2) It generates an embedding that forms a part of the input for the force estimation head as follows:
\vspace{-6pt}
 \begin{equation}
	\label{4}
	\begin{split}
	 GAP(\hat{O}_{btn}^{sn}) = \frac{1}{h\times w}\sum_h \sum_w \hat{O}_{btn}^{sn}[h,w] 
	\end{split}
\end{equation} 

\noindent 

Where (1) is the global average pooling that generates $\vec v^{sn}_{btn}[n]$ in which $v$ is an embedding of length $n$ obtained from the feature maps $\hat{O}$ with $n$ channels in the bottleneck ($btn$) of sub-network $sn$. Furthermore, $\hat O_{btn}^{sn}$ is input to the first block of the decoder. The decoder consists of four blocks, each upscaling the input by a factor of 2. In every block $b$ of sub-network $sn$, a transposed convolution increases the input feature maps' size, which are then concatenated with $\hat{O}^{sn,b}_{copy}$, received from the corresponding encoder block. This layer is followed by 2 successive 2D convolution layers with kernels of $3\times3$ and a ReLu activation function. $\hat{O}^{b,sn}_{dec}$ is the output of block $b$ generated by the second convolution in sub-network $sn$. In both sub-network, the above-mentioned output feeds the next block as well as supplies a part of input embedding for the estimation head:
\vspace{-3pt}
 \begin{equation}
	\label{4}
	\begin{split}
	\vec v^{b, sn}_{dec}[n] = GAP(\hat{O}^{b,sn}_{dec}) 
	\end{split}
\end{equation}

\noindent where $\vec v$ is the embedding of decoder's block $b$ in sub-network $sn$ produced by the the global average pooling (1). The fourth block of the decoder in each sub-network reconstructs the feature maps to match the size of the input images. Afterwards, the output is passed through a 2D convolution layer with a single $1 \times 1$ kernel and a sigmoid activation function. This layer maps the feature maps onto the segmentation plane $\hat O^{sn}$. In fact, the segmented catheter from the X-Ray images is achieved by solving a classification problem in the output layer of each sub-network through the following loss function: 
\vspace{-3pt}
\begin{multline}
\label{7}
L^{sn}_{seg}(\hat O^{sn}, t_{sn}) = \\
- \frac{1}{N} \sum_i^{N} t_{sn}^i log (\hat O^{sn}_i) + (1 - t_{sn}^i) log(1 - \hat O^{sn}_i)
\end{multline}

\noindent the equation above is a binary cross entropy and $t$ denotes the label map for the corresponding of sub-network $sn$ and $N$ is the total number pixels.
As previously discussed, the regression head receives input embeddings from the bottlenecks $\vec v^{sn=1}_{btn}$ and $\vec v^{sn=2}_{btn}$, as well as feature vectors transferred by every block of the decoder in each sub-network, denoted as $\vec{v}^{b, sn}_{dec}$ (refer to equation (1) and (2)). 
To this end, a part of the final input can be created as follows: 
\vspace{-3pt}
\begin{align}
\label{7}
\vec{V}_{sn} &= \vec v^{sn}_{btn} \oplus [ \|_1^b \vec{v}^{b, sn}_{dec} ]
\end{align}

\noindent In equation (4), '$\|$' represents a series of concatenations of the feature vectors generated by block $b=1$ to $b=4$ (as mentioned in equation (2) of the decoder in each sub-network, denoted as $sn$. The resultant vector is then concatenated ($\oplus$) with $\vec{v}^{sn}_{btn}$. To clarify, the feature vector from the bottlenecks reflects the global properties of the input image. Moreover, the embeddings from the decoder infuse the input image's attributes, while simultaneously working to eliminate elements other than the catheter's shape. This enables the estimation head to focus on variations in the catheter's shape, thereby modeling the applied forces without being distracted by noise. This process results in the creation of a general vector $\vec{V}_{sn}$ for each sub-network. Given these vectors, the input to the regression head is formed as follows:
\vspace{-3pt}
\begin{align}
\label{7}
\vec{V}_{reg} &= \vec{V}_{sn=1} \oplus \vec{V}_{sn=2}
\end{align}

\noindent here $\vec{V}_{reg}$
incorporates the aforementioned traits from two distinct angles, thereby providing the regression head with stereo vision capabilities. This input vector traverses through two hidden dense layers with 64 and 32 units, respectively. The activation function used is ReLu except for the output layer. It generates $\hat{O}^{reg}$ as a 3D force vector through 3 units with a linear activation function. The force estimation head solves a regression problem by optimizing a Mean Squared Error (MSE) loss function as follows: 
\vspace{-3pt}
\begin{equation}
\label{7}
L_{reg}(\hat{O}^{reg}, t_{reg}) = \frac{\sum_{k=1}^d (\hat{O}^{reg}_k- t_{reg}^k)^2}{d}
\end{equation} 
\noindent Here, $t$ represents the target force vector, which consists of $d=3$ elements corresponding to the forces along the $x$, $y$, and $z$ directions. As explained, H-Net is an end-to-end architecture comprising two sub-networks, each with a segmentation head, and includes a force estimation head for the entire network. Consequently, the total loss function for the network is defined as follows:

\vspace{-10pt}
\begin{equation}
\label{7}
L_{total} = \beta_1 L^{sn=1}_{seg} + \beta_2 L^{sn=2}_{seg} + \beta_3 L_{reg} 
\end{equation}

\noindent in this context, $L^{sn=1,2}_{seg}$ is derived from equation (3), and $L_{reg}$ is obtained from equation (6). Additionally, $\beta$ adjusts the contribution weight of each loss function within the total loss.
The aforementioned total loss function, denoted by equation (7), is optimized using the Root Mean Squared Propagation (RMSprop) optimizer, which tunes the parameters of the model. It is worth mentioning that all layers and their parameters in both the encoder and decoder are shared between the two sub-networks, with the goal of maintaining the model's computational complexity and memory utilization at manageable levels.
%

\vspace{-15pt}
\section{Evaluation and Discussion}
This section describes the data preparation and model configuration for the training and validation phases. It also provides a benchmark for evaluating the performance of H-Net in comparison to other solutions found in the literature. The topics mentioned above will be covered in two subsections. Subsection A will first review the data preparation process in detail. This will be followed by a comprehensive explanation of the model's configuration for training and inference. Lastly, the model's performance will be evaluated for both the segmentation and force estimation heads by comparing it with existing solutions in the literature in Subsection B.
\begin{figure}[t]
	
	\centering
	\includegraphics[trim={0mm 0mm 0 0},clip, width=0.4\textwidth]{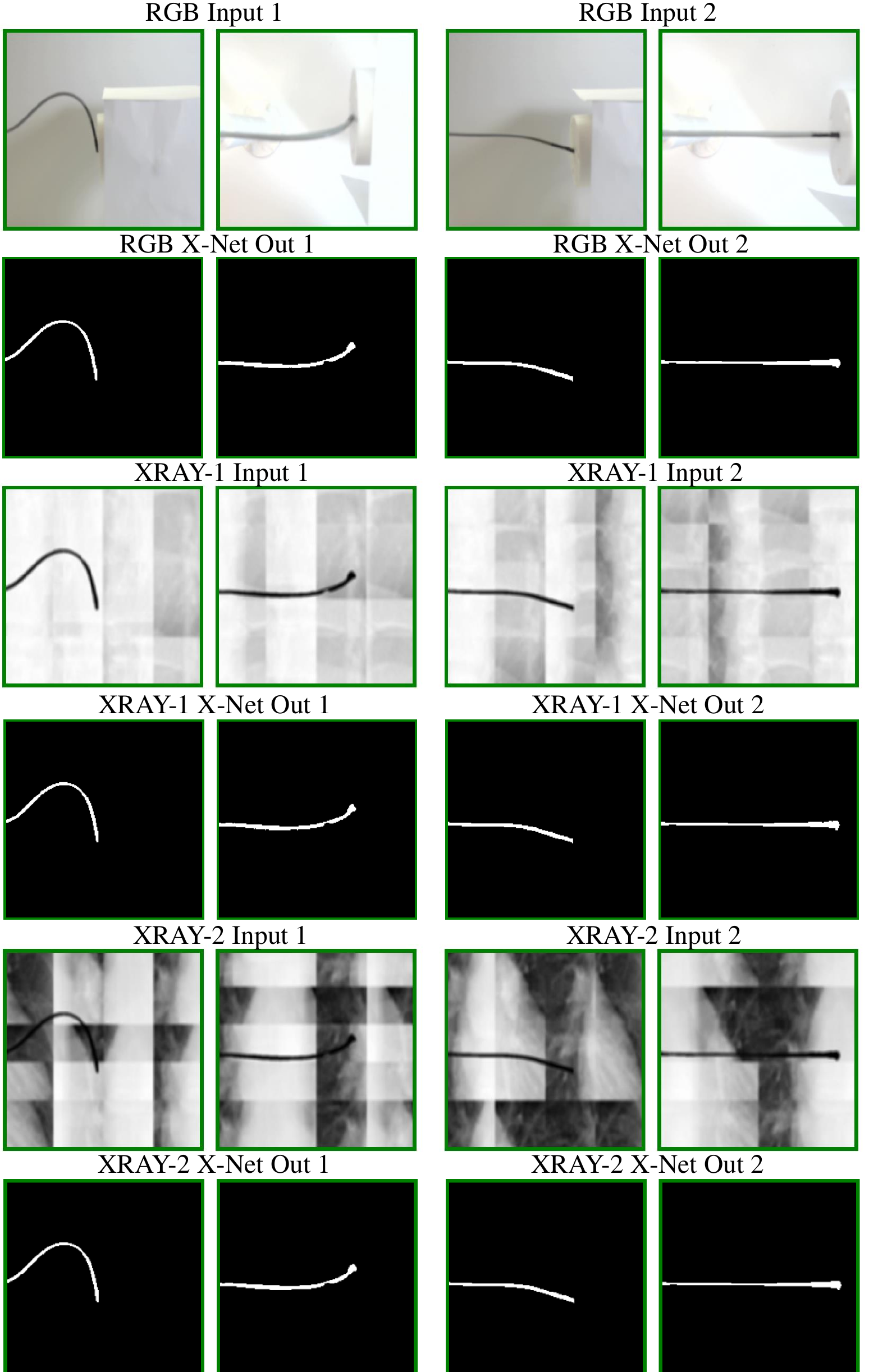}
	\caption{The diagram demonstrates two samples from RGB, synthetic XRay-1 and XRay-2 dataset. The outputs of H-Net for each sample are shown as well. For each dataset, H-Net was already trained on its train set. }
	\label{fig:net}
\end{figure}
\vspace{-13pt}
\subsection{Dataset Preparation and Model Configuration}
H-Net utilizes three types of data, each divided into training, testing, and validation sets. The first data type includes RGB images of a catheter, previously used in the Y-Net study \cite{khodashenas2021, fekri2021deep, y-net}. The second and third types consist of synthetic X-Ray images. As outlined in subsection II.A, the difficulty level of synthetic X-Ray images for the model is determined by the background regions, which could affect the model's ability to differentiate between the catheter and the background. Examples of these datasets are illustrated in Fig. 3. The XRay-1 samples comprise synthetic images with a lower difficulty level, characterized by more consistent intensity levels in the selected background regions. Conversely, the XRay-2 samples includes synthetic images featuring a mix of dark and bright pixels, randomly distributed across the background mesh.
Each of the datasets described above contains 19,500 samples. Every sample includes two images, each with dimensions of $224 \times 224 \times 3$, which are normalized before being inputted into the network. The targets for the segmentation head consist of two segmentation maps, also with dimensions of $224 \times 224 \times 1$, where each pixel indicates either the catheter or the background. For the force estimation head, the target for each sample is a force vector along the $x$, $y$, and $z$ axes. 
Furthermore, each dataset was shuffled and divided into training, validation, and test sets. The training set, comprising $80\%$ of the data or 13,650 samples, was the largest portion. The remaining $20\%$ of the data was equally split between the test and validation sets, with each containing 2,925 samples.
\par The H-Net processes two input images at a time, which are each fed into their respective sub-network's encoders. These encoders comprise four blocks. Each block consists of two successive 2D convolutional (conv) layers with ReLU activation functions, followed by a max-pooling layer. The 2D conv layers are equipped with 32 filters of size $3\times3$ and have a stride of 1. The max-pooling layers feature a kernel size of $2\times2$, with a stride of 2. The feature maps produced by these encoders are then processed by the two bottleneck blocks in both sub-networks. These bottleneck blocks are similar to the encoder blocks but are equipped with 2D conv layers without the max-pooling layer. 
In each sub-network of H-Net, the decoder receives the output from the encoder. It upscales the feature maps through four blocks to generate segmentation maps for each input image at the segmentation (classification) heads. Each block within the decoder includes a 2D convolution transpose layer, followed by two convolution layers. Both the 2D convolution and 2D convolution transpose layers are outfitted with 32 filters, each sized $3\times3$. The convolution layers have a stride of 1, whereas the convolution transpose layers use a stride of 2. The final block of each decoder functions as the classification head and includes an additional 2D convolution layer. This layer, equipped with a single filter of size $1\times1$, is responsible for producing the segmentation map. 
As mentioned in sub-section II.B, the H-Net estimated the applied forces through a regression head in which it receives the input from both the bottlenecks as well as the decoders. This head processes the input embeddings using three stacked dense layers, with 64, 32, and 3 units, respectively. The first two layers employ a ReLU activation function, while the final layer uses a linear activation function to output the estimated forces. The weights of each head's loss ($\beta$) are equal. The configured H-Net was trained on the compiled training set using batches of 32 samples, with a learning rate of $1 \times 10^{-4}$, over 100 epochs. To prevent overfitting, the model's performance was monitored by evaluating it on the validation set at the end of each epoch, facilitating the implementation of an early-stopping method.
\vspace{-15pt}
\subsection{Results and Discussions}
\vspace{-5pt}
Given that H-Net produces two segmentation maps for a catheter from two different angles, as well as the 3D forces, the performance of these two tasks will be evaluated separately. In assessing H-Net's capability to estimate 3D forces, the model's performance is evaluated using the test set as unseen data. As previously mentioned, H-Net was trained on three distinct datasets: RGB, synthetic XRay-1, and synthetic XRay-2. To this end, the following metrics are used in order to report the performance: Mean Absolute Errors ($\textbf{MAE}$), Mean Squared Errors $(\textbf{MSE})$, Root Mean Squared Errors ($\textbf{RMSE}$), $\mathit{\mathbf{R^2}}$ and the ratio of RMSE (R) and the average of the maximum (M) forces in the available directions ($\mathit{\mathbf{R/M}}$).
Table I presents a comparison of H-Net's force estimation performance with the state-of-the-art learning-based (Y-Net) force estimators from the literature, utilizing the aforementioned metrics. Y-Net utilizes a similar dataset generated by the same setup. However, as discussed in Subsection II.A, in this study, the synthetic data was derived from the base dataset previously used in Y-Net. 
\begin{center}
\begin{table}[t]
	\caption{ The table reports the performance of the H-Net comparing with the learning-based method in the literature.}
	\label{table}
	\centering
	\begin{tabular}{c c c c c c c}
	\hline
	     Model & Inp/FD & MSE & MAE & RMSE & $R^2$ & R/M   \\
		\hline \hline \\
		
		Y-Net \cite{y-net}
            & Seg/3
		& 2.8e-05 
		& 0.0039
		& 0.005
		& 0.98
		& 0.026
        \\

		
		\hline
		\\
		H-Net 
            & RGB/3
		& 3.6e-05 
		& 0.0039
		& 0.005
		& 0.98
		& 0.026
        \\
		\hline
		\\
		H-Net 
            & XRay1/3
		& 3.3e-05 
		& 0.0039
		& 0.0049
		& 0.98
		& 0.026
        \\
		\hline
		\\
		H-Net
            & XRay2/3
		& 4.3e-05 
		& 0.0046
		& 0.0058
		& 0.97
		& 0.03
        \\
		\hline
  
		\hline\hline
	\end{tabular}
\end{table}
\end{center}
\vspace{-26pt}
\vspace{0pt}
\begin{figure}[b]
	\centering
        \captionsetup{skip=0pt}
	\includegraphics[trim={0mm 0mm 0 0},clip, width=0.48\textwidth]{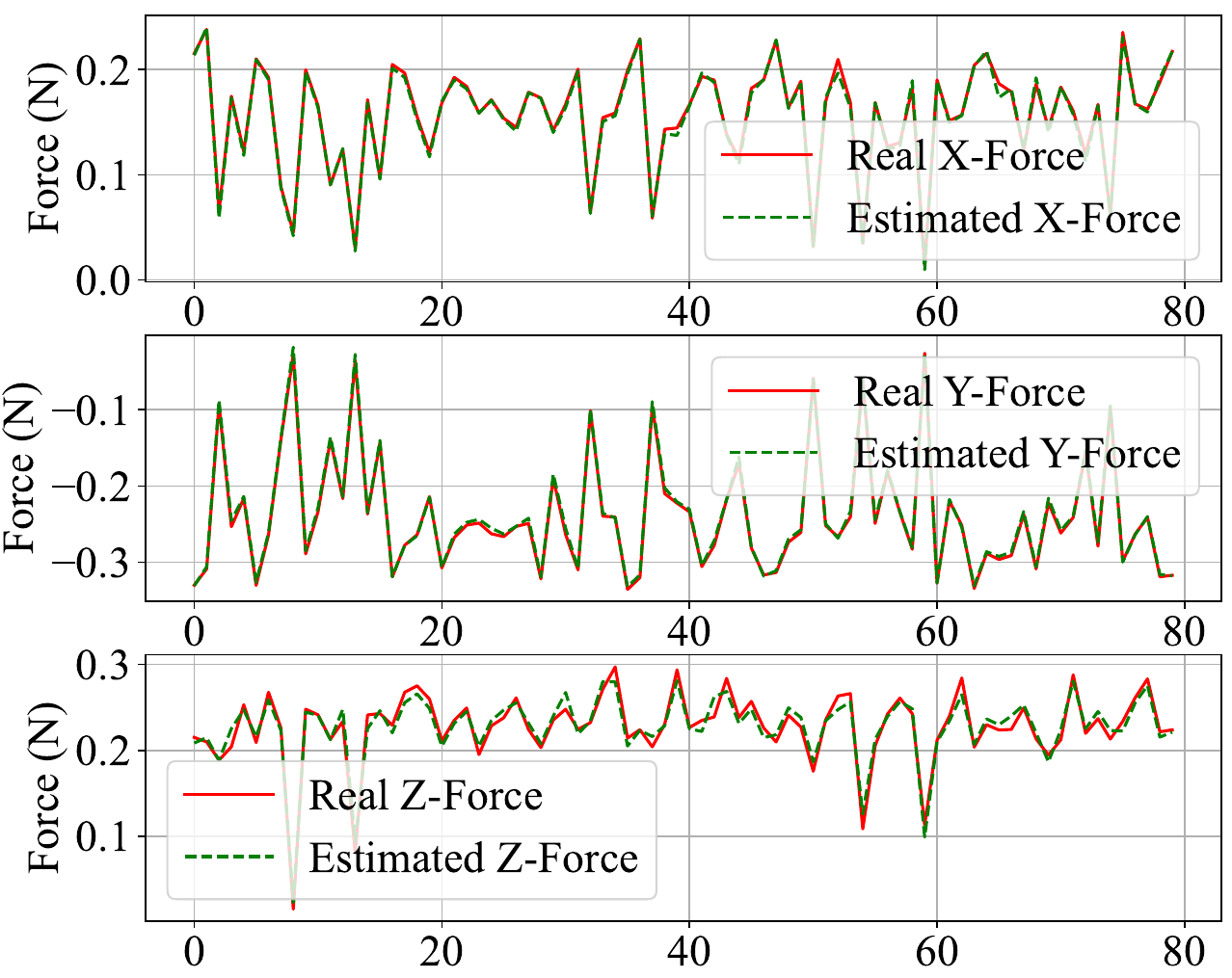}
	\caption{The diagram demonstrates the predicted forces in $x$, $y$ and $z$, given 80 input samples of XRay-2's test set.}
	\label{fig:samplepred}
\end{figure}

\begin{figure*}[t]
    \centering
    \begin{subfigure}{.30\textwidth}
        \centering
        \includegraphics[width=\linewidth]{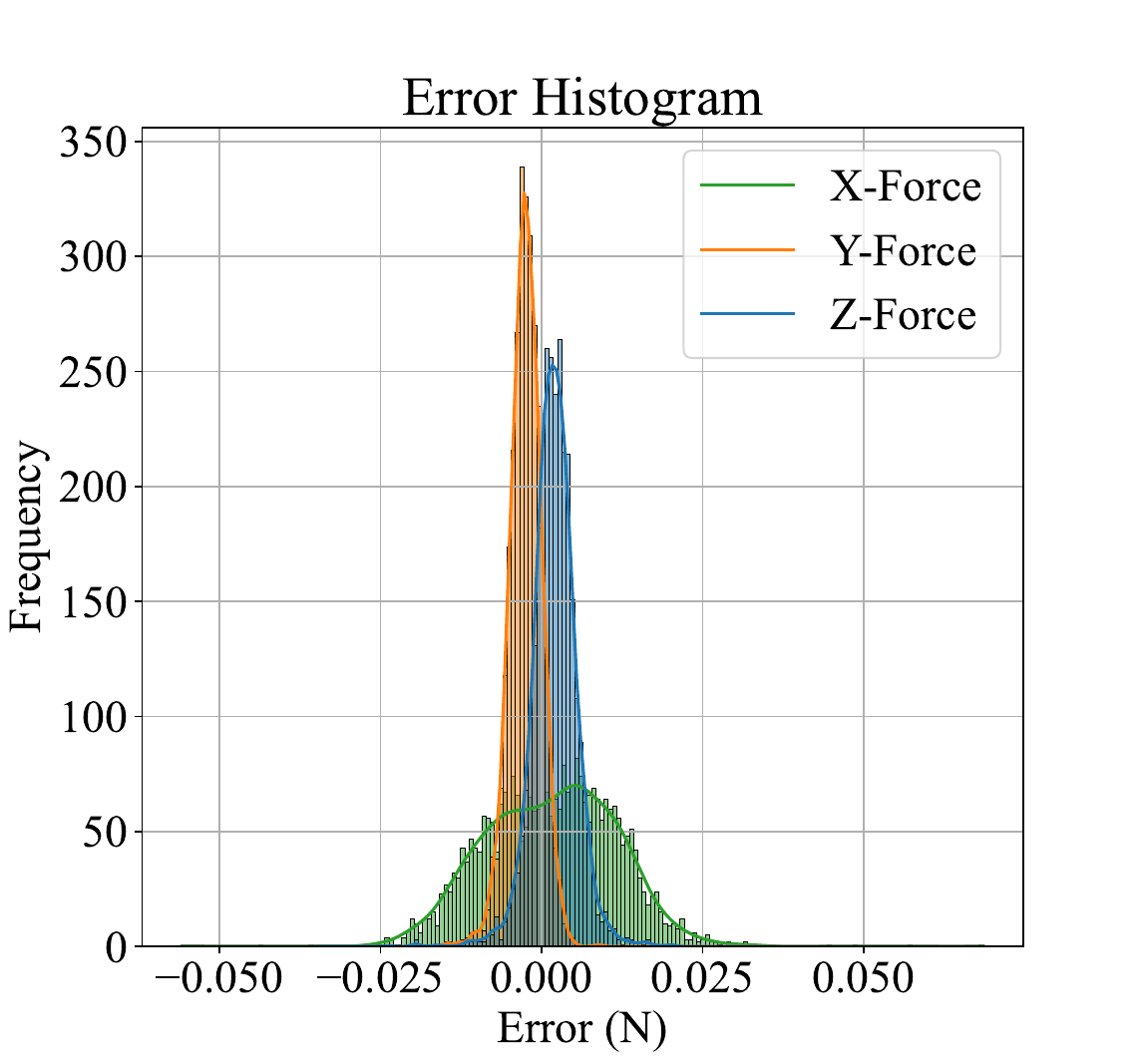}
        \caption{H-Net tested on XRay-2}
        \label{fig:sub1}
    \end{subfigure}%
    \begin{subfigure}{.30\textwidth}
        \centering
        \includegraphics[width=\linewidth]{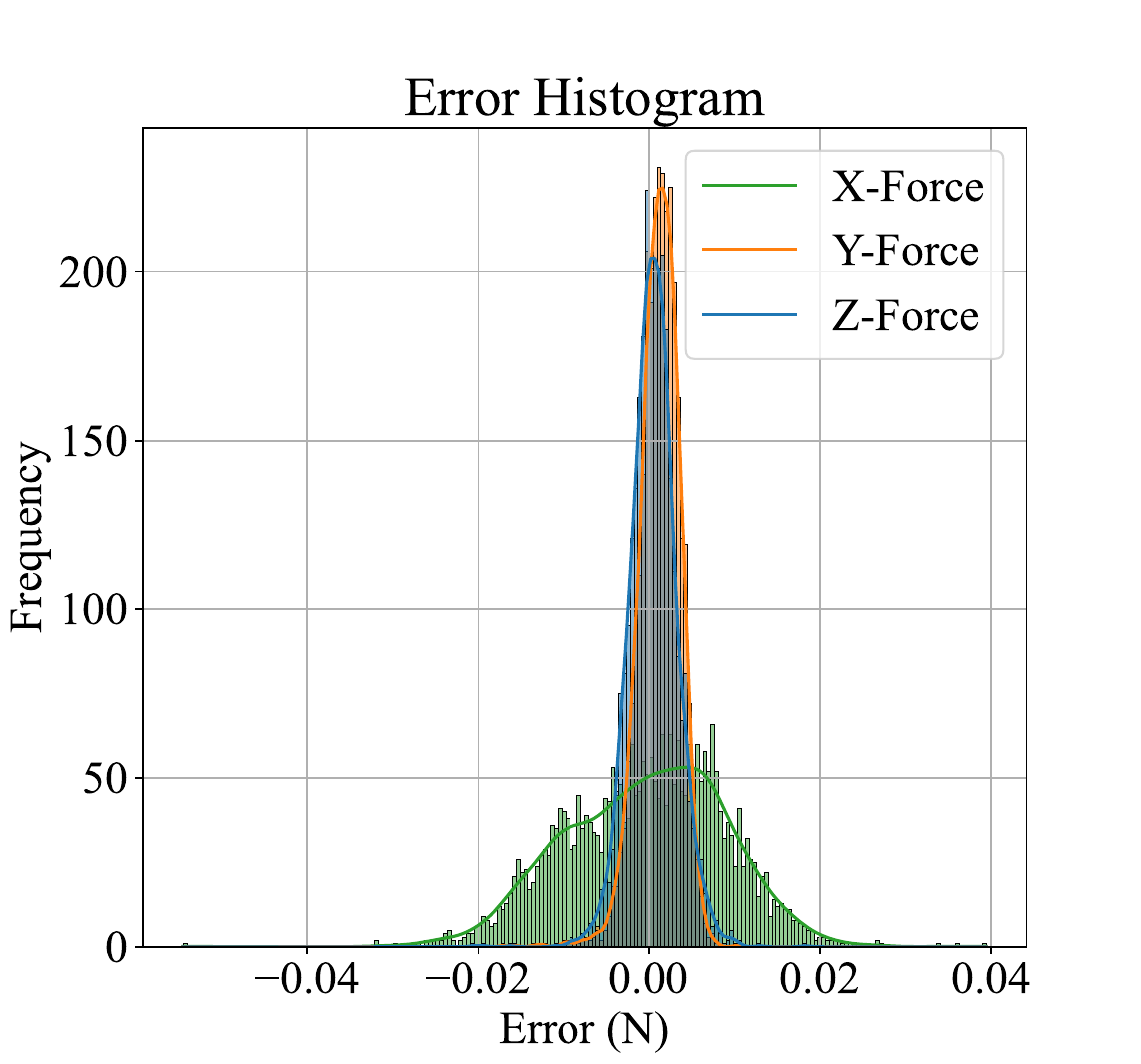}
        \caption{H-Net tested on XRay-1}
        \label{fig:sub2}
    \end{subfigure}%
    \begin{subfigure}{.30\textwidth}
        \centering
        \includegraphics[width=\linewidth]{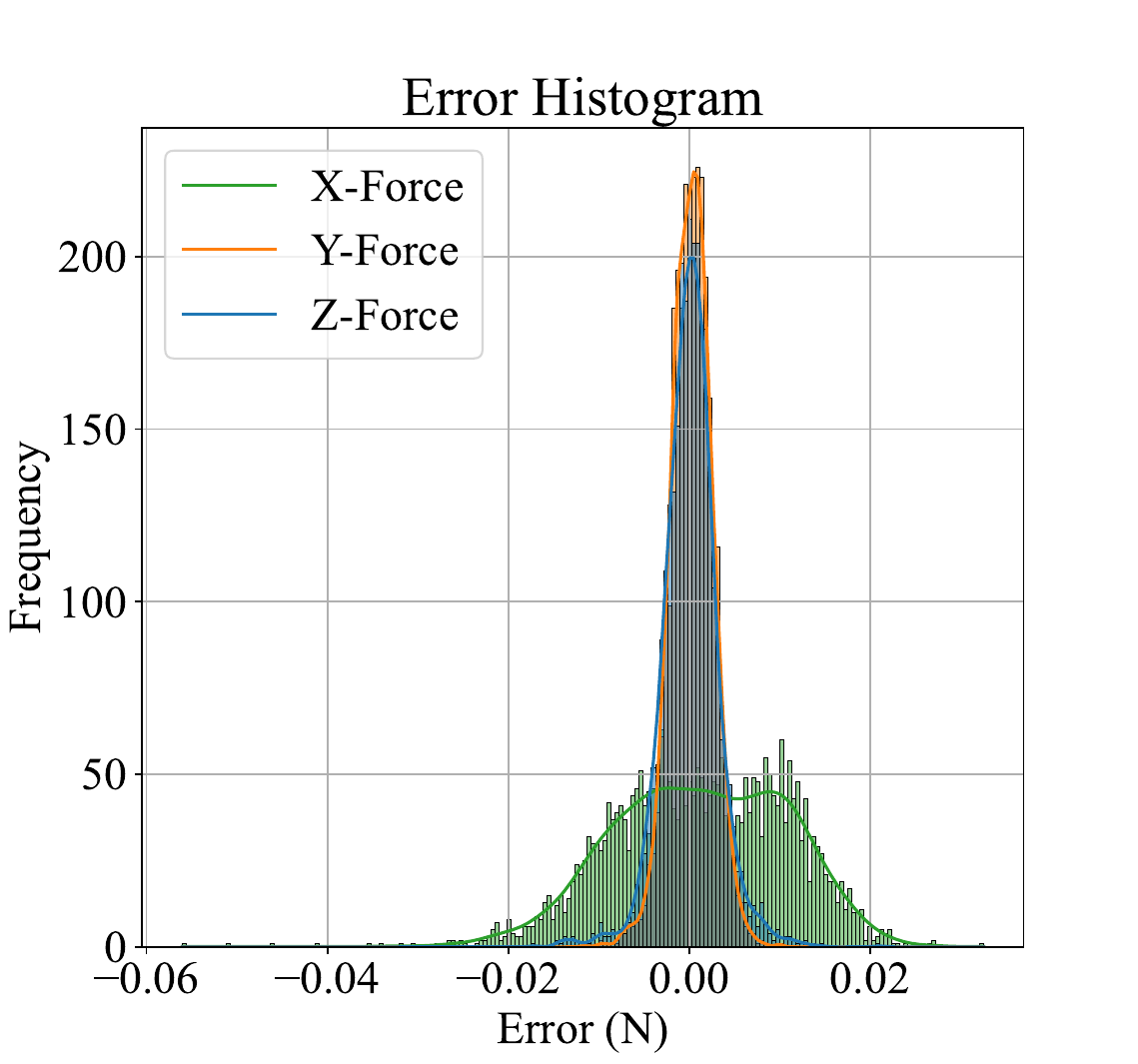}
        \caption{H-Net tested on RGB}
        \label{fig:sub3}
    \end{subfigure}

    \begin{subfigure}{.30\textwidth}
        \centering
        \includegraphics[width=\linewidth]{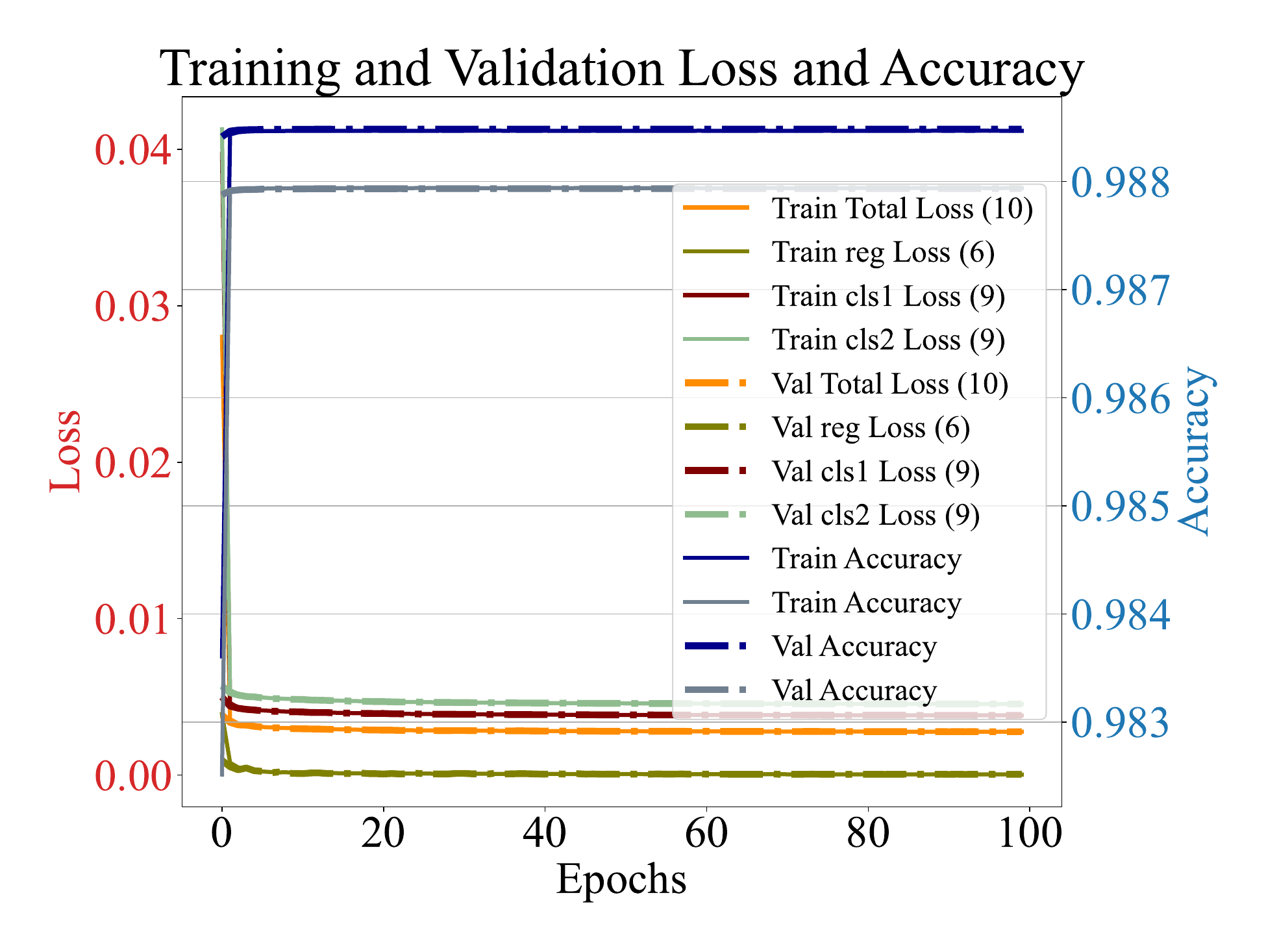}
        \caption{H-Net tested on XRay-2}
        \label{fig:sub4}
    \end{subfigure}%
    \begin{subfigure}{.30\textwidth}
        \centering
        \includegraphics[width=\linewidth]{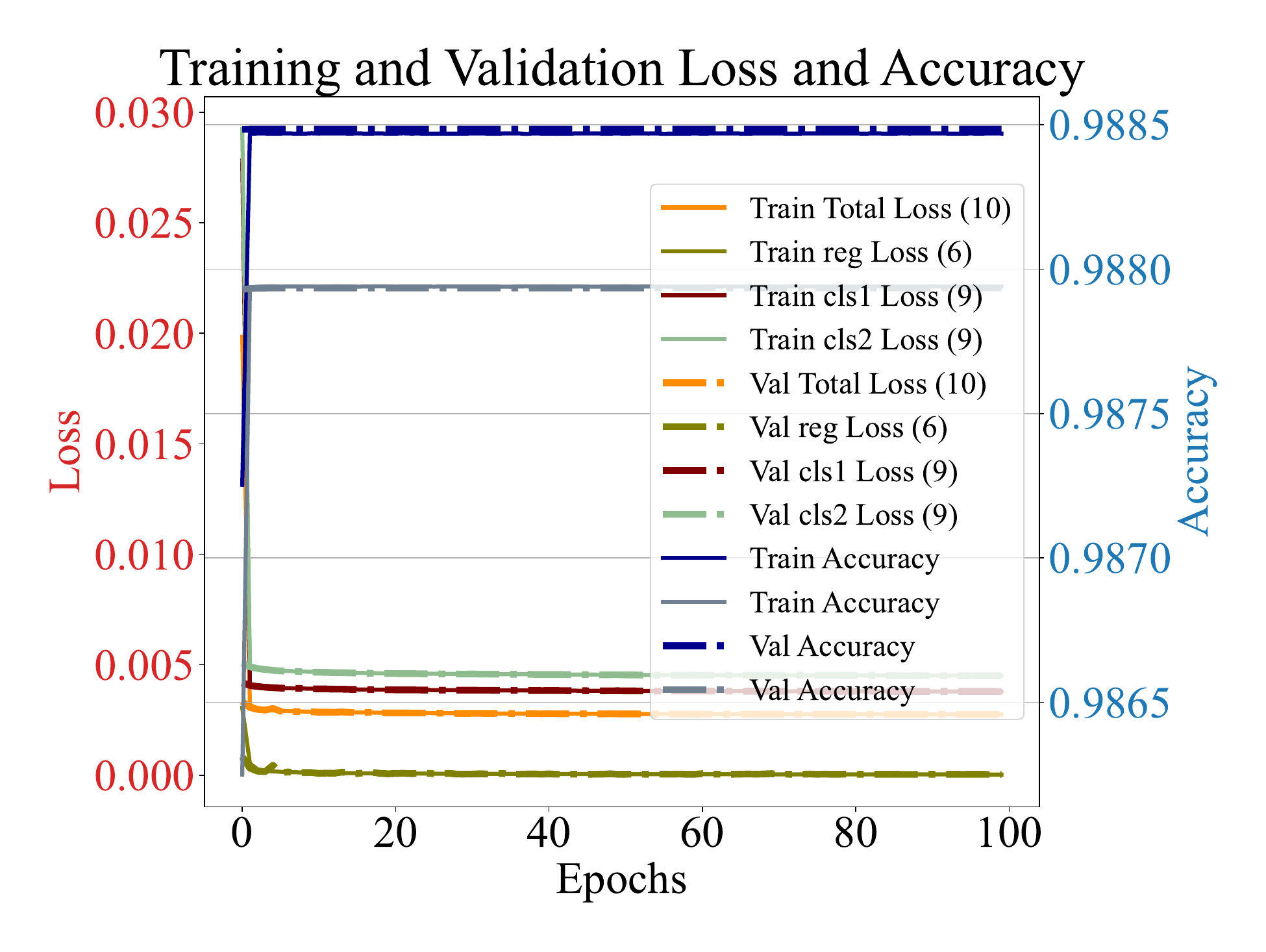}
        \caption{H-Net tested on XRay-1}
        \label{fig:sub5}
    \end{subfigure}%
    \begin{subfigure}{.30\textwidth}
        \centering
        \includegraphics[width=\linewidth]{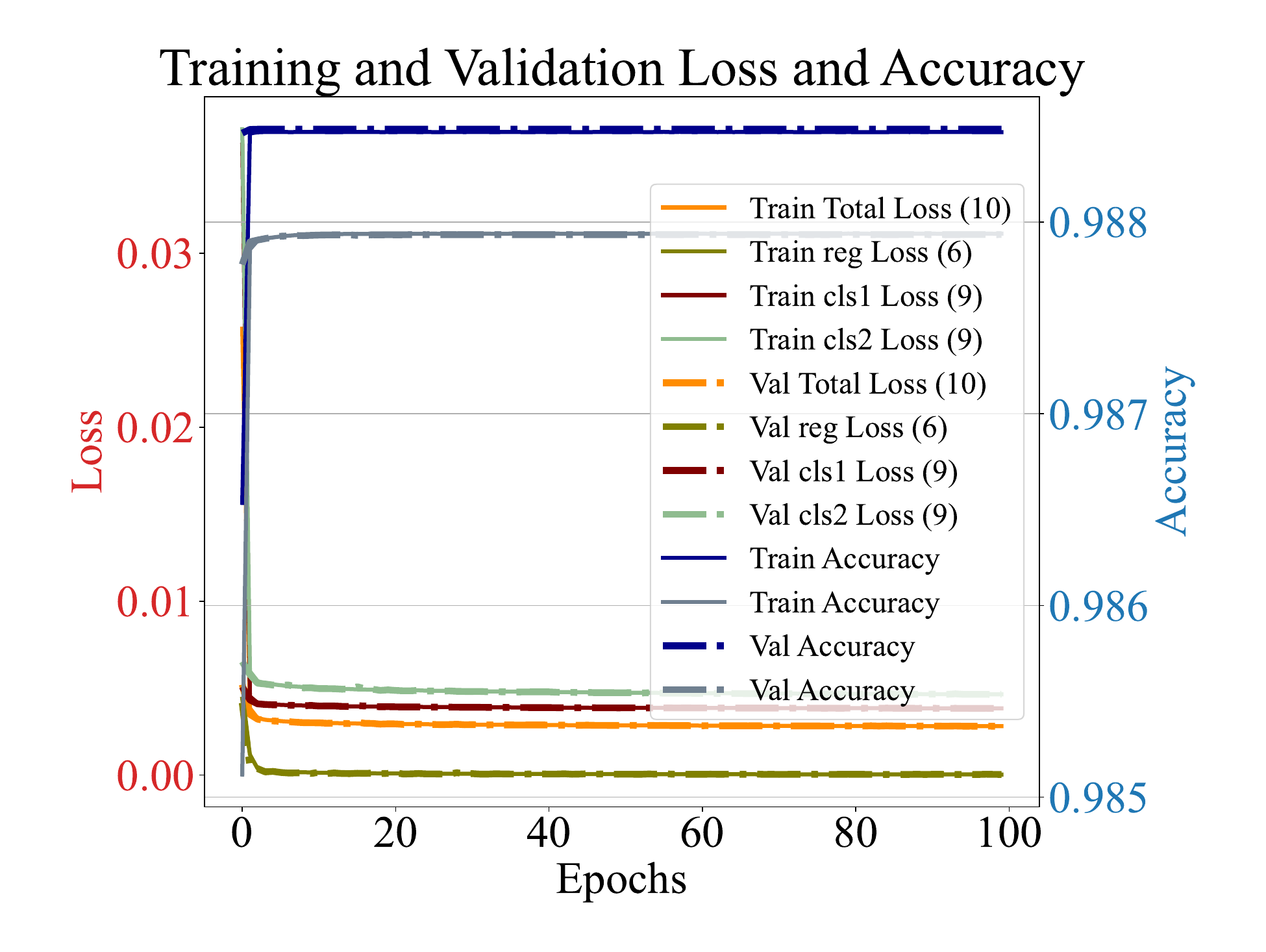}
        \caption{H-Net tested on RGB}
        \label{fig:sub6}
    \end{subfigure}

    \caption{The diagram plots the histogram of errors for the output of H-Net's force estimation head on each dataset in subplot (a) to (c). It also figures the training and validation loss and accuracy of H-Net in subplot (d) to (f). }
    \label{fig:test}
\end{figure*}
\vspace{-1pt}
Y-Net is a CNN-based architecture that addresses a significant challenge in learning-based force estimation by introducing 3D force estimation capabilities to the network \cite{y-net}. Force Dimension (FD) in the table refers to the dimension of output forces along $x, y$ and $z$ (FD = 3). Similar to H-Net, it processes two images of a catheter simultaneously, visualizing the catheter from two perspectives. This enables the model to also capture variations along the $z$ axis. However, Y-Net still is fed by segmented catheter images as inputs. Both H-Net and Y-Net estimate force in 3D, providing a fair basis for comparison between the two models. When comparing H-Net trained on RGB data with Y-Net (three layers/blocks) trained on manually segmented (seg) images, the results across all metrics indicate a relatively similar performance. 
\par 
The impact of synthetic X-ray images on the performance of H-Net was investigated using the XRay-1 and XRay-2 datasets. Compared to the RGB set, the model achieved similar results on the synthetic set with a smoother background (XRay-1). The MAE error of H-Net increased by approximately 17.9\% when exposed to the more challenging synthetic XRay-2 dataset, compared to its performance on the XRay-1 and RGB datasets. However, the force estimation remains precise, aligning with state-of-the-art results. Fig. 4 displays the estimated forces alongside their corresponding actual forces for 80 samples from the XRay-2 test set, presented in the x, y, and z. Obviously, every samples contains two images that are inputted into the model during the inference phase. The histogram of force errors for H-Net is depicted in Fig. 5. Comparing subplot (a) with subplots (b) and (c), it is noticeable that the error distribution for estimated forces along the z-axis for XRay-2 is more widely spread, while the errors for the other force axes are not as well centered around 0. As suggested in \cite{y-net}, since the desired mean error value is above zero, a mechanism can evaluate the model's error at various camera positions relative to the catheter to identify the optimal position with the minimum prediction error. Additionally, compiling data from a rotating camera can also address this issue. With that in mind, H-Net is the only network that processes raw input images and accurately predicts forces in 3D across all datasets.
This demonstrates its capability as a force estimator, particularly in extracting variations in the catheter's deflection from noisy images captured from two perspectives. 
\par
The segmentation heads of H-Net were separately evaluated on the test sets using accuracy (\textbf{acc}), the mean Intersection over Union (\textbf{mIoU}) metric and the mean Dice (\textbf{mDice}). Additionally, the number of trainable parameters for each model is reported in millions. This metric indicates the computational complexity of the model. Table II serves as a benchmark, comparing each head of H-Net with three commonly used semantic segmentation models in medical applications: FCN, U-Net and Hr-Net \cite{fcn,FCN-cat, unet, unet-2024_cat, hrnet, hrnet_cat}. 
In this work, FCN and U-Net were randomly initialized and trained on the synthetic XRay-1 and XRay-2 datasets. Meanwhile, the HR-Net, pre-trained on Cityscapes, was fine-tuned on these datasets. 
Since all models in the benchmark, except for H-Net, have only one head, they have been trained and evaluated using images from both angles through one head. 
Although FCN contains the highest number of trainable parameters compared to the other models in the benchmark, the mIoU for this model is lower across all datasets. Hr-Net is the second smallest model in the benchmark. This model is able to precisely find the catheter's region in all datasets. Compared to H-Net, the mIoU for Hr-Net is slightly higher. However, H-Net is significantly smaller in size. Additionally, the improved mIoU in Hr-Net could be attributed to the model being fine-tuned on a previously trained on Cityscape dataset. U-Net encompasses 34 million parameters, but its mDice and mIoU are approximately identical to those of H-Net, which achieves similar performance with a much smaller set of parameters. 
\par H-Net is the only architecture in the benchmark capable of segmenting two images simultaneously.
The mIoU and Dice reported for H-Net are the averages of the results from both heads.
Although H-Net has two encoders, two decoders, and a force estimation head, the network's total number of trainable parameters is the lowest in the benchmark, thanks to its shared layers.
To achieve a low parameter count while ensuring high performance, we implemented filter pruning, removed normalization layers (batch and layer normalization), and designed the model with shared parameters across tasks. Unlike general-purpose segmentation models that require extensive features to handle diverse image content, our model is dedicated exclusively to catheter segmentation and 3D force estimation. This domain-specific focus allowed us to significantly reduce the number of parameters, as our architecture is optimized solely for catheter-related features rather than general object variability. By employing a shared encoder-decoder structure for both tasks, we further minimized parameter redundancy, learning a unified feature representation that supports both tasks efficiently. Together, these strategies, i.e. filter pruning, the removal of normalization layers, and domain-specific design allowed us to develop a lightweight, high-performing model with under 500K parameters. This provides additional benefits, including lower memory usage, faster model training, easier model loading on GPUs with smaller memory capacities, improved model generalization, and reduced risk of overfitting.

Fig. 3 illustrates the outputs of the segmentation heads for H-Net, showcasing two samples from the RGB, XRay-1, and XRay-2 datasets. It is worth noting that the accuracy for all models in the benchmark remains almost constant, while the mIoU and Dice varies. This variation can be attributed to the binary classification nature of the dataset, which is highly imbalanced. This imbalance arises due to the number of pixels representing the catheter is significantly smaller than those representing the background. Also, the Dice coefficient generally gives a higher score than IoU for the same overlap level due to its sensitivity to the intersection, especially when mask sizes are similar. Finally, the second row of Fig. 5 displays the training and validation loss and accuracy of H-Net across all datasets throughout 100 epochs with no sign of overfitting. Cls and reg refer to the classification (segmentation) and regression (force estimation) heads, respectively. The implementation, training and validation of all models in the benchmark were conducted on an Ubuntu 20.04 machine with an NVIDIA 2080 GPU.

\begin{center}
\begin{table}[t]
	\caption{ The table reports the performance of the H-Net's segmentation heads comparing with the literature.}
	\label{table}
	\centering
	\begin{tabular}{c c c c c }
	\hline
	     models / (datasets) & params (m) & acc & mIoU & mDice \\
		\hline \hline \\
		FCN  (rgb)\cite{fcn}
		& 134
            & 99.2
		& 94.0
		& 96.8
	    \\
     \hline \\
		FCN  (xray-1)\cite{fcn}
		& 134
        & 99.2
		& 94.1
		& 96.8
	    \\
     \hline \\
		FCN  (xray-2)\cite{fcn}
		& 134
        & 99.1
		& 93.8
		& 96.7
	    \\
		\hline
		\\
		U-Net (rgb)\cite{unet}
		& 34
            & 99.8
		& 95.7
		& 98.0

		\\
		\hline

		\\
		U-Net (xray-1)\cite{unet}
		& 34
            & 99.7
		& 95.7
		& 98.0

		\\
  \hline
		\\
		U-Net (xray-2)\cite{unet}
		& 34
            & 99.7
		& 95.4
		& 97.8

		\\
  \hline
  \\
		Hr-Net (rgb) \cite{hrnet}
		& 9.636
            & 98.8
		& 96.1
		& 98.4

        \\
		
		\hline
		\\
  Hr-Net (xray-1) \cite{hrnet}
		& 9.636
            & 98.8
		& 96.1
		& 98.5

        \\
		
		\hline
		\\
  Hr-Net (xray-2) \cite{hrnet}
		& 9.636
            & 98.8
		& 95.8
		& 98.0
		
        \\
        \hline
		\\
		H-Net (rgb)
		& 0.460
            & 98.8
            & 95.7
		& 98.0
        \\
        \hline
		\\
		H-Net (xray-1)
		& 0.460
            & 98.8
            & 95.7
		& 98.0
        \\
        \hline
		\\
		H-Net (xray-2)
		& 0.460
            & 98.8
            & 95.5
		& 97.8
        \\
		
		\hline\hline
	\end{tabular}
\end{table}
\end{center}
\vspace{-45pt}
\section{Conclusion}
In this work, we proposed a novel multitask, multi-input, multi-output, encoder-decoder architecture for 3D force estimation and semantic segmentation in cardiac ablation. Additionally, we developed a synthetic X-Ray generator to compile a dataset that closely resembles real X-Ray images obtained from a biplane fluoroscopy machine. Given two input images showing the catheter from two angles, the network processes them simultaneously. It accurately segments the catheter in both images and estimates the applied forces in 3D. All aforementioned tasks are accomplished within an end-to-end architecture. The proposed network was trained and tested on an RGB dataset and two synthetic X-Ray datasets, each with different levels of difficulty. Both the segmentation and force estimation tasks, performed by the three heads of the network, were compared with several existing force estimators and semantic segmentation models. The proposed model demonstrated state-of-the-art performance. This is the first time an end-to-end architecture has been capable of segmenting the catheter and estimating forces in 3D. For future work, the network could be revised to incorporate the impact of the catheter's tendon on force estimation.

\vspace{-10pt}

\bibliographystyle{IEEEtran}
\bibliography{ref}









\vfill


\end{document}